\documentclass[a4paper,showpacs,prl,twocolumn]{revtex4}
\usepackage{amsfonts}
\usepackage{graphicx}
\usepackage{color}
\usepackage{amsmath}
\usepackage{amssymb}
\usepackage{latexsym}
\usepackage{psfrag}

\begin{document}

\title{Effect of Electron Interaction on Statistics of Conductance
Oscillations in Open Quantum Dots: Does the Dephasing Time Saturate? }
\author{S. Ihnatsenka}
\affiliation{Solid State Electronics, Department of Science and Technology (ITN), Link%
\"{o}ping University, 60174 Norrk\"{o}ping, Sweden}
\date{\today }
\author{I. V. Zozoulenko}
\affiliation{Solid State Electronics, Department of Science and Technology (ITN), Link%
\"{o}ping University, 60174 Norrk\"{o}ping, Sweden}
\date{\today }

\begin{abstract}
We perform self-consistent quantum transport calculations in open quantum
dots taking into account the effect of electron interaction. We
demonstrative that in the regime of the ultralow temperatures $2\pi
k_BT\lesssim\Delta$ ($\Delta $ being the mean level spacing), the electron
interaction strongly affects the conductance oscillations and their
statistics leading to a drastic deviation from the corresponding predictions
for noninteracting electrons. In particular, it causes smearing of
conductance oscillations, which is similar to the effect of temperature or
inelastic scattering. For $2\pi k_BT\gtrsim\Delta$ the influence of electron
interaction on the conductance becomes strongly diminished. Our calculations
(that are free from phenomenological parameters of the theory) are in good
quantitative agreement with the observed ultralow temperature statistics
(Huibers \textit{et al.}, Phys. Rev. Lett. \textbf{81}, 1917 (1998)). Our
findings question a conventional interpretation of the ultralow temperature
saturation of the coherence time in open dots which is based on the
noninteracting theories where the electron interaction is neglected and the
agrement with the experiment is achieved by introducing additional
phenomenological channels of dephasing.
\end{abstract}

\maketitle

\textit{Introduction.} Decoherence of quantum states due to interaction with
an environment represents one of the fundamental phenomena in quantum
physics. In low-dimensional semiconductor structures such as quantum dots,
wires, antidots the decoherence processes are central to electronic
transport and spin/charge manipulation. A temperature dependence of the
phase coherence time in open quantum dots $\tau _{\varphi }$ has been a
focus of significant experimental activity during the past decade\cite%
{Marcus,Bird,Clarke,Huibers,Huibers_b,Huibers_c,Pivin,Hackens}. The
experiments show that $\tau _{\varphi }$ increases as the temperature $T$
decreases following a dependence $\tau _{\varphi }\sim T^{-\gamma }$, where $%
1\lesssim \gamma \lesssim 2$. This behavior indicates that several
mechanisms might be simultaneously responsible for the electron decoherence,
including electron scattering with a large energy transfer\cite{e-e
scattering} (leading to $\tau _{\varphi }\sim T^{-2}$ dependence), and a
small energy-transfer (Nyquist) scattering\cite{Altshuler} (giving a rate $%
\tau _{\varphi }\sim T^{-1}$). Surprisingly, practically all experiments
report a remarkable effect of a saturation of the phase coherence time at
ultralow temperatures $T\lesssim 100$ mK. The origin of this effect is not
understood and at present no theory is available addressing the electron
decoherence in confined ballistic systems. It should be noted that a similar
effect of a saturation of the phase coherence time is also found in
nanoscaled metallic wires and the origin of this saturations also remains
open and highly debated\cite{metallic wires}.

An experimental determination of the dephasing time $\tau _{\varphi }$ is
typically based on predictions of the random matrix theory (RMT)\cite{RMT}
for the statistics for quantum transport such as the mean and variance of
the conductance oscillations, the weak localization corrections, the
probability distribution of conductance and others\cite%
{RMT,statistics_inelastic,statistics_RMT}. The RMT is essentially
noninteracting theory relying on a one-electron description of quantum
transport. In order to fit the experimental data, the dephasing time $\tau
_{\varphi }$ is included as a phenomenological parameter of the theory
typically within a B\"{u}ttiker's fictitious voltage probe or as an
imaginary potential in the Hamiltonian\cite{RMT,statistics_inelastic}. A
deviation of the experimental data from the predictions of a purely coherent
model of noninteracting electrons is then attributed to inelastic scattering
due to dephasing which is extracted using $\tau _{\varphi }$ as a fitting
parameter.

How does the electron interaction affect the conductance oscillations in the
open dots? This question was posed in several theoretical studies with
somehow conflicting conclusions\cite%
{Brouwer_1999,Brouwer_2005,Indlekofer,open_dot}. For example, Brouwer and
Aleiner\cite{Brouwer_1999} argued that the Coulomb interactions enhance the
weak localization and increase conductance fluctuations, whereas Brouwer
\textit{et al.}\cite{Brouwer_2005} questioned these conclusions. None of the
above studies however addressed the problem of the low-temperature
saturation of the coherence time. In the present paper we, based on the
first-principle self-consistent quantum transport calculations, study the
effect of the electron interaction on the probability distribution of the
conductance, $P(G)$, in open dots. We demonstrate that for sufficiently high
temperatures (when the transport energy window exceeds the mean level
spacing, $2\pi k_{B}T\gtrsim \Delta $, the corresponding distributions $P(G)$
for interacting and noninteracting electrons are practically the same.
However, in the opposite limit of ultralow temperatures, $2\pi
k_{B}T\lesssim \Delta $, the distributions of $P(G)$ are strikingly
different for noninteracting and interacting electrons. We compare our
calculated statistics for interacting electrons with corresponding
experimental results of Huiberts \textit{et al.} \cite{Huibers_b} and find a
good quantitative agreement. Our results therefore strongly indicate that a
deviation of the experimental data from the RMT predictions in the regime of
ultralow temperatures can be accounted for by the electron interaction alone
without introducing additional channels of the inelastic scattering. Our
findings thus question the conclusion concerning the saturation of the the $%
\tau _{\varphi }$ in open dots which is obtained neglecting electron
interaction and under the assumption that the above deviation is due to the
inelastic scattering only.

\textit{Model}. We consider an open quantum dot defined by split-gates in
GaAs heterostructure, see Fig. 1. The Hamiltonian of the whole system (the
dot + the semi-infinite leads) can be written in the form $H=H_{0}+V(\mathbf{%
r}),$ where $H_{0}=-\frac{\hbar ^{2}}{2m^{\ast }}\left\{ \left( \frac{%
\partial }{\partial x}-\frac{eiBy}{\hbar }\right) ^{2}+\frac{\partial ^{2}}{%
\partial y^{2}}\right\} $ is the kinetic energy in the Landau gauge, and the
total confining potential $V(\mathbf{r})=V_{conf}(\mathbf{r})+V_{H}(\mathbf{r%
})$ is the sum of the electrostatic confinement (including contributions
from the top gates, the donor layer and the Schottky barrier), and the
Hartree potential, see \cite{open_dot,QPC} for details
\begin{equation}
V_{H}(\mathbf{r})=\frac{e^{2}}{4\pi \varepsilon _{0}\varepsilon _{r}}\int d%
\mathbf{r}\,^{\prime }n(\mathbf{r}^{\prime })\left( \frac{1}{|\mathbf{r}-%
\mathbf{r}^{\prime }|}-\frac{1}{\sqrt{|\mathbf{r}-\mathbf{r}^{\prime
}|^{2}+4b^{2}}}\right) ,  \label{V_H}
\end{equation}%
where $n(\mathbf{r})$ is the electron density, the second term corresponds
to the mirror charges situated at the distance $b$ from the surface, and $%
\varepsilon _{r}=12.9$ is the dielectric constant of GaAs. Note that the dot
and the leads are treated on the same footing, e.g. the Coulomb interaction
and the magnetic field are included both in the lead and in the dot regions.
We consider the spinless electrons because in relatively large dots as those
studied here the electrons are spin degenerate\cite{Folk,Martin}. We also
neglect the exchange and correlation effects, which have been shown to
affect the calculated conductance only marginally\cite{open_dot}.

To outline the role of the electron interaction in the conductance of open
quantum dots we also calculate magnetoconductance in the Thomas-Fermi (TF)
approximation. In this approximation the self-consistent electron density is
given by the standard TF equation,$\frac{\pi \hbar ^{2}}{m^{\ast }}%
n(x,y)+V_{conf}(r)+V_{H}(r)=E_{F}$. The electron density and the total
confining potential calculated within the TF approximation do not capture
quantum-mechanical quantization of the electron motion and the resonant
energy structure in the dot. The utilization of the TF approximation for
modeling of the magnetotransport in open system is therefore conceptually
equivalent to noninteracting one-electron transport calculations, where,
however, the total confinement is given by a smooth realistic potential.

The magnetoconductance through the quantum dot in the linear response regime
is given by the Landauer formula, $G=-\frac{2e^{2}}{h}\int dE\,T(E)\frac{%
\partial f_{FD}\left( E-E_{F}\right) }{\partial E}$. A detailed description
of the self-consistent conductance calculation (as well as the validity and
applicability of the method and the Hamiltonian) are given in our previous
publications \cite{open_dot,QPC}. Note that the present approach corresponds
to the ``first principle" magnetoconductance calculation (within the
effective mass approximation), that starts from a geometrical layout of the
device, is free from phenomenological parameters and not relying on model
Hamiltonians whose validity is poorly controlled.

\textit{Results and discussions.}
\begin{figure}[tb]
\includegraphics[scale=0.9]{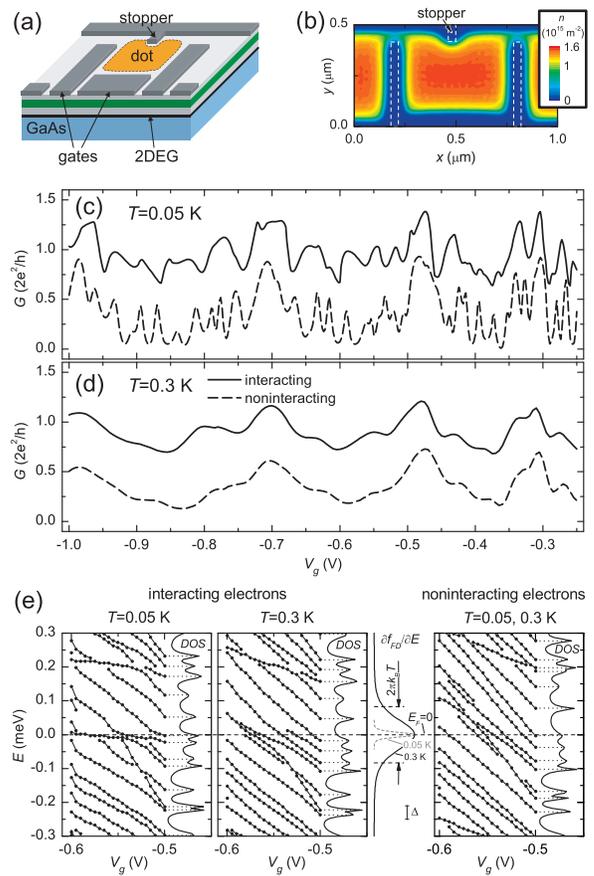} 
\caption{(Color online). (a) A schematic layout of a split-gate open quantum
dot defined in a GaAs heterostructure. The geometrical size of the dot is $%
660\times 520$nm, the width of the leads is 540nm. The widths of the cap,
donor and spacer layers are 14 nm, 36 nm and 10 nm respectively), the donor
concentration is $0.6\cdot 10^{24}$ m$^{-3}$ . (b) A representative
self-consistent electron density in the dot (note that the densities for the
interacting (Hartree) and noninteracting (TF) electrons are not
distinguishable on the scale of the figure). White dashed lines indicate a
geometry of the metallic gates. (c),(d) The calculated conductance of the
dot for interacting (Hartree) and noninteracting (TF) electrons as a
function of the gate voltage $V_{g}$ for (a) $T=50$mK and (b) $T=300$mK; $%
B=20$mT. [The conductance curves for interacting electrons are shifted by $%
e^{2}/h$.] (e) Resonant energy structure (i.e. positions of the peaks in the
DOS as a function of the $V_{g}$) for different temperatures; A calculated
DOS is shown for $V_{g}=-0.5V$; the Fermi energy is set $E_{F}=0$. [Note
that the resonant energy structure for noninteracting electrons is
practically undistinguishable for the given temperatures]. Inset shows the
derivative of the Fermi-Dirac distribution function for $T=0.05$K and 0.3K
and the transport window $2\protect\pi k_{B}T$.}
\label{fig1}
\end{figure}
\begin{figure}[tb]
\includegraphics[scale=1.1]{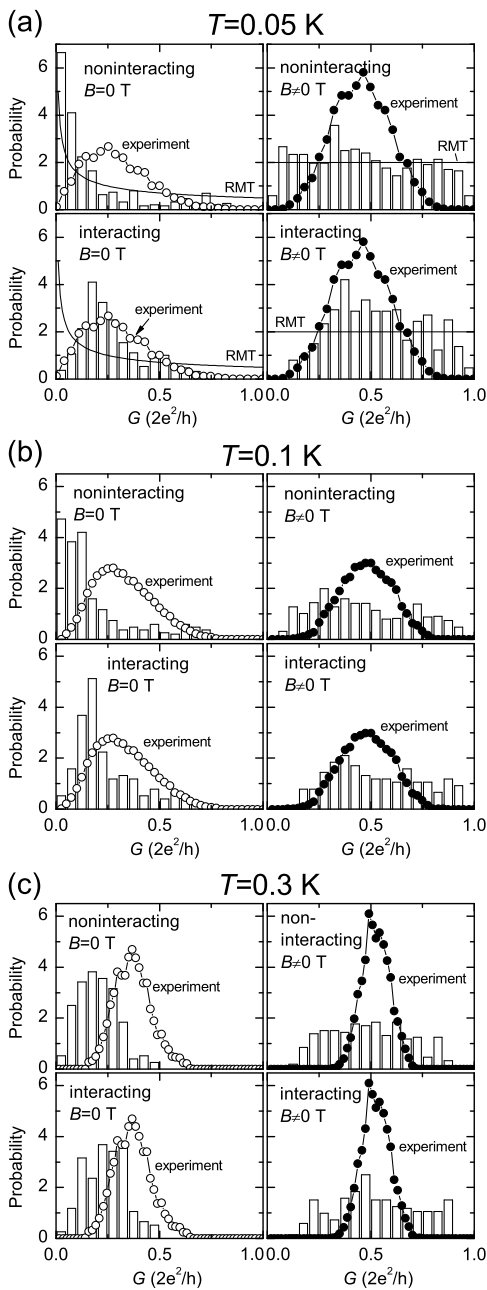} 
\caption{Probability distribution of the conductance $P(G)$ for interacting
and noninteracting electrons for different temperatures for the cases of the
time-reversed symmetry ($\protect\beta =1$) and the broken time-reversed
symmetry ($\protect\beta =2$). The experimental data is adapted from Ref.
\protect\cite{Huibers_b}. Solid lines in (a) corresponds to the predictions
of the RMT ($T=0$, no dephasing).}
\label{fig2}
\end{figure}
Figure \ref{fig1}(c) shows the conductance of the open quantum dot
calculated in the Hartree and TF approximations (interacting and
noninteracting electrons respectively) for $T=50$mK. The parameters of the
dot are indicated in Fig. \ref{fig1} and are chosen close to those studied
experimentally by Huibers \textit{at al.} \cite{Huibers_b}. All the results
discussed in this paper correspond to one propagating mode in the quantum
point contact openings. The striking difference between the conductance
curves is clearly manifested in a strong suppression of the high frequency
components of the oscillations for the interacting electrons in comparison
to the noninteracting case. \textit{Thus, the electron interaction causes an
apparent smearing of the conductance oscillations, which is similar to the
effect of the temperature or inelastic scattering}. This smearing of
oscillations is caused by the pinning of resonant levels to the Fermi energy
in the vicinity of resonances\cite{open_dot}. This is illustrated in Fig. %
\ref{fig1} (e) which shows an evolution of the peak position of the resonant energy
levels. In the vicinity of the resonances the DOS of the dot is enhanced such that
electrons with the energies close to $E_{F}$ can easily screen the external potential.
This leads to the \textquotedblleft metallic" behavior of the system when the electron
density in the dot can be easily redistributed to keep the potential constant. As a
result, in the vicinity of a resonance the system only weakly responds to the external
perturbation (change of a gate voltage, magnetic field, etc.), i.e. the resonant levels
becomes pinned to the Fermi energy (see Ref. \cite{open_dot} for a detailed discussion of
the pinning effect). For noninteracting electrons the nonlinear screening and hence the
pinning effect are absent, such that the successive dot states sweep past the Fermi level
in a linear fashion, see Fig. \ref{fig1} (e).

The pinning of resonant levels drastically affects the conductance
probability distribution $P(G)$. Figure \ref{fig2}(a) shows $P(G)$
calculated for interacting and noninteracting electrons for the cases of a
time-reversal symmetry, $\beta=1$ ($B=0$) and a broken time-reversal
symmetry, $\beta=2$, ($B\neq 0$) for $T=50$mK. The time-reversal symmetry is
broken by application of a magnetic field $B\gtrsim \phi_0/A$, where $%
\phi_0=h/e$ is the flux quantum and $A$ is the dot area (typically, $B\sim
20-40$mT). Figure \ref{fig2} (a) shows that the statistics of the
conductance distribution $P(G)$ for the case of noninteracting electrons
closely follows the corresponding RMT predictions for $\tau _{\varphi }=0$
and $T=0$\cite{RMT,statistics_RMT} both for $\beta=1$ and $\beta=2$. At the
same time, the statistics for the interacting electrons are strikingly
different from those for the noninteracting case. Thus, \textit{due to the
effect of the electron interaction, the ultra-low temperature statistics of
the conductance oscillations of quantum dots are not described by the RMT}.

As the temperature increases, the difference between the conductances $%
G=G(V_{g})$ as well as between the corresponding conductance distributions $%
P(G)$ for interacting and noninteracting electrons diminishes, see Fig. \ref%
{fig2} (b) ($T=100$mK). For sufficiently high temperature this difference
disappears, see Figs. \ref{fig1} (d), \ref{fig2} (c) ($T=300$mK). The reason
for that is that the temperature strongly reduces the effect of resonant
level pinning. Indeed, when the transport energy window, $\sim 2\pi k_{B}T$,
(determined by the condition when the derivative of the Fermi-Dirac
distribution is distinct from zero, see Fig. \ref{fig1} (e)) exceeds the
mean level spacing $\Delta =\frac{2\pi \hbar ^{2}}{m\ast A}$ ($A$ being the
dot area) 
the conductance is mediated by several levels. As a result, several levels
always contribute to screening at the same time and hence the screening
efficiency of the dot is affected very little when a gate voltage or a
magnetic field are varied. A quenching of the pinning for temperatures $2\pi
k_{B}T\gtrsim \Delta $ due to suppression of the resonant level screening is
illustrated in Fig. \ref{fig1} (e) ($T=300$mK). Note that for the dot under
consideration the condition $2\pi k_{B}T=\Delta $ corresponds to $T\approx
100$mK. Thus, \textit{for }$2\pi k_{B}T\gtrsim \Delta $\textit{\ the effect
of electron interaction on the conductance is strongly suppressed such that
the conductance and their probability distributions for interacting and
noninteracting electrons are practically the same.}

Let us now compare our results with available experimental data. The
probability distribution $P(G)$ in open quantum dots with one propagating
channel in the leads was studied by Huiberts \textit{et al.} \cite{Huibers_b}%
. Figure \ref{fig2} (a) shows that in the regime of the ultralow temperatures, $T=50$mK,
the calculated conductance statistics for interacting electrons agree quite well with the
corresponding experimental distribution $P(G)$ both for $\beta =1$ and $\beta =2$. The
measured conductance distribution $P(G)$ in Ref. \onlinecite{Huibers_b} was well
described by the RMT predictions where the inelastic scattering was introduced using
$\tau _{\varphi }$ as a fitting parameter. Our results, instead, demonstrate, that once
the electron interaction is accounted for, the agreement with the experiment for $%
2\pi k_{B}T\lesssim \Delta $ is achieved without assuming additional inelastic scattering
channels. We thus conclude that for the regime of ultralow temperatures the
experimentally inferred value of $\tau _{\varphi }$ might be greatly underestimated which
implies that the dephasing time does not saturate. As the temperature increases, the
calculated conductance distribution starts to deviated from the experimental statistics,
see Fig \ref{fig2} (b),(c). As discussed above, for the temperature $2\pi k_{B}T\gtrsim
\Delta $ the electron interaction practically does not affect the conductance
oscillations and their statistics. Thus, for $2\pi k_{B}T\gtrsim \Delta $ the difference
between the calculated and experimental statistics can be attributed to the effect of
dephasing. Our criterium for the transition temperature $2\pi k_{B}T\sim \Delta $ is
consistent with the findings reported by Bird \textit{et al.} \cite{Bird,Pivin} and
Clarke \textit{et al.}\cite{Clarke} who find a
saturation behavior of $\tau _{\varphi }$ at transition temperatures $%
T_{onset}$ near the mean-level spacing. A relation between $T_{onset}$ and $%
\Delta $ was also discussed by Hackens \textit{et al.} \cite{Hackens}.
However, some experiments\cite{Huibers_c} do not show a clear relation
between $T_{onset}$ and $\Delta $, such that more systematic studies are
needed in order to prove the connection between $T_{onset}$ and $\Delta $.

We stress that our calculations are performed for purely coherent electrons.
The dephasing effects can be easily included in our model phenomenologically
through an imaginary potential in the Hamiltonian\cite{RMT}. We do not
provide a systematic fit of the experiment simply because of a computation
burden related to this task: each point on the conductance plot requires up
to one hour of a processor time. We however note that such a fit is outside
the scope of our study, where we focuss on the role of the electron
interaction in a regime of the ultralow temperatures, $2\pi k_{B}T\lesssim
\Delta $.

The findings reported in this paper outline importance of the
\textquotedblleft first-principle" self-consistent quantum transport
calculations for open quantum dots. Indeed, accounting for both global
electrostatics through the Hartree potential, Eq. (\ref{V_H}), and the
quantum mechanical quantization in a self-consistent way is essential for
revealing of the pinning effect that causes a drastic difference in the
conductance of the interacting and noninteracting electrons. Note that this
effect would not be captured in the approach utilizing model Hamiltonians
(like those of Ref. \cite{Brouwer_1999,Brouwer_2005} where the electron
interaction is accounted thought the classical capacitance charging).

To conclude, we demonstrate that for ultralow temperatures $2\pi
k_{B}T\lesssim \Delta $ the electron interaction drastically changes the
statistics of the conductance oscillations in open dots leading to a
significant departure from the conventional RMT description of
noninteracting electrons. Our results demonstrate that the deviation of the
observed statistics at ultralow temperatures from the RMT predictions can be
accounted for by the electron interaction alone, such that a conclusion of
the dephasing time saturation based on noninteracting electron picture
should be revised.

\textit{Acknowledgement.} We thank J. P. Bird for discussion and valuable comments.

\end{document}